\documentclass[twocolumn,prb]{revtex4}

\usepackage{graphicx}
\usepackage{dcolumn}
\usepackage{bm}

\def\spose#1{\hbox to 0pt{#1\hss}}
\def\lta{\mathrel{\spose{\lower 3pt\hbox{$\mathchar"218$}}
     \raise 2.0pt\hbox{$\mathchar"13C$}}}
\def\gta{\mathrel{\spose{\lower 3pt\hbox{$\mathchar"218$}}
     \raise 2.0pt\hbox{$\mathchar"13E$}}}

\begin{document}


\title[Friction in Stellar Systems]{Dynamical Friction in Stellar Systems: an introduction}

\author{H\'ector Aceves}
\email{aceves@astrosen.unam.mx}
\affiliation{
Instituto de Astronom\'{\i}a, UNAM. Apartado Postal 877, Ensenada, B.C. 22800,
M\'exico. 
}%

\author{Mar\'{\i}a Colosimo}
 \email{mcolosim@exa.unicen.edu.ar}
\affiliation{
Facultad de Ciencias, 
Universidad Nacional del Centro de la Provincia de Buenos Aires. 
Tandil, Argentina}

\date{\today}

\begin{abstract}
An introductory exposition of Chandrasekhar's gravitational 
\emph{dynamical friction}, appropriate for an undergraduate class in mechanics,
  is presented. This friction results when
a massive particle moving through a ``sea'' of much lighter 
star particles experiences a retarding force du to an exchange of energy and momentum. 
 General features
of dynamical friction are presented, both in an elementary and in a more
elaborate way using hyperbolic two-body interactions.
  The orbital decay of a massive particle in an
  homogeneous gravitational system is solved analytically, that leads   
  to an underdamped harmonic oscillator type of motion.
 A numerical integration of the equation of motion in a more realistic case
 is done. These results are compared to those of an $N$-body computer
 simulation. 
  Several problems and projects are suggested to students for further study.
\end{abstract}


\pacs{Valid PACS appear here}

\maketitle

\section{\label{sec:intro} Introduction} 

Classical mechanics, perhaps the oldest of the physical sciences, continues
to be an area of intensive research, both in its
foundations\cite{Marsden,Hestenes} and applications,\cite{Solar,Diacu} and
a source of discussion and examples in teaching.
Applications range from the modeling of cellular mechanical
processes\cite{cell} to solar system dynamics\cite{Solar} and galactic
systems.\cite{BT,Saslaw,Aarseth}

In describing nature students learn from their first courses, and
 particularly in laboratory experiments,  
that ``the forces on a single thing already involve approximation, and if we
have a system of discourse about the real world, then that system, at least
for the present day, must involve approximations of some
kind''; as mentioned by Feynman on introducing  the subject of friction.\cite{Feynman1}

 This phenomenon 
is usually introduced in text-books\cite{AF,HRK,BPC1,French}
 and lectures by considering the  
 slide of a material block on a surface, and a distinction between 
static and kinetic friction is made. 
 A classical example of the effect of a 
  friction-like force is the motion of a mass attached to a spring inside a
  viscous medium, where the corresponding differential equation is solved, and
  its behavior studied. 
  At the end, one invariably needs to state
 that friction and its origin is a complicated matter, involving complex
 interactions at the atomic and molecular level among the surfaces in
 contact.\cite{Palmer,FrictionRL01,FrictionRL02,Ringlein}

Several non-typical examples of mechanical friction for introductory courses
 exist,\cite{Parkyn,Lapidus,Molina,Simbach} that help  both teachers and students alike in lectures on mechanics. 
 All friction related  
 problems are a background for discussing the important  
 connection between 
 the work-energy theorem and dissipative systems.\cite{Sherwood,Mallin,Arons}

 The purpose of this paper is to bring an example from 
astronomy\cite{Sagan,Arny,Shu} 
  closely related to standard mechanical friction, namely: {\sl dynamical
 friction}. This process was first introduced  in stellar systems by
  Subrahmanyan Chandrasekhar.\cite{Chandra1943,ChandraBook} In brief,
 a massive particle
  $m$ experiences a drag force when moving in a ``sea'' of much lighter 
 star particles $m_*$ by exchanging  energy and momentum. An 
 elementary 
 understanding requires only some basic ideas from  mechanics, and hence
 suitable for presentation in introductory courses.

 Dynamical friction is important in astronomical studies of, for example: the fate of galaxy
 satellites\cite{W89,VW,FFM05} or globular clusters\cite{McMillanPZ03} orbiting their host galaxies, the substructure of dark halos
 surrounding galaxies,\cite{vdBosch99,Zhao04,BullockJ05} and the motion of
 black holes in the centers of galaxies.\cite{kim04}  It
 has been proposed to explain the formation of binaries in the
 Kuiper-belt\cite{Goldreich02}, and the migration of Jupiter-mass
 planets in other solar systems from the outer parts where they presumably
 formed ($\gta 1$~AU) to the small orbital distances ($\lta 0.1$~AU) at which
 they are observed.\cite{DelPopolo} It even has been considered  in the
 motion of cosmic strings.\cite{Avelino95}

The presentation of this topic to students, in a lower or
upper-undergraduate class on mechanics\cite{Taylor,Kibble} or computational
physics,\cite{Spencer-CP}  will enhance their
appreciation of physics in describing nature and expose them to another 
example of classical mechanics.
 Furthermore,  students will obtain a glimpse of an  area of
astronomical research important for the understanding of the fate and behavior
of stellar systems.

 The organization
 of this paper is as follows. In Section~II  basic elements of the theory of  dynamical
 friction are presented. Firstly, elementary arguments are used to elucidate
 them. Secondly, Chandrasekhar's approximation
 using two-body hyperbolic Keplerian collisions is considered.
 In Section~III a simple analytical problem for the motion of a massive
 particle in an ideal homogeneous stellar system is solved; a damped harmonic
 oscillator is found. In Section~IV
 a more realistic astronomical example that requires the numerical
 integration of the equation of motion is presented. Comparison with a computer experiment
 is done afterwards. Final comments as well as some ideas for problems and
 projects of further study are  provided in Section~IV. An appendix contains
 some astronomical units and standard units used in
 gravitational computer simulations.

\section{Dynamical Friction}\label{sec:df}

Two equivalent approaches to compute the dynamical friction  a massive
particle  $m$ experiences as it moves through a stellar system of much lighter
stars $m_*$ are the
following.\cite{BT}  
(1) Particle $m$ produces a region of star overdensity behind it, much like
the wake behind the motion of a ship, that in turn exerts a gravitational pull
on $m$ leading to its deceleration.\cite{Mulder} (2) Particle $m$
moves in the ``sea'' of lighter particles $m_*$ and an energy exchange occurs, 
increasing that of the lighter ones at the expense of the heavy one leading
to a breaking force for $m$. In the latter picture the basic features of
dynamical friction are easier to compute and understand by elementary methods.
 Here the latter picture is taken. 

\subsection{Elementary estimate}\label{sec:df1}

\begin{figure}[!t]
\includegraphics[width=8cm]{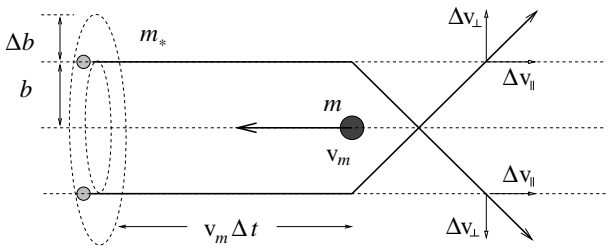}
\caption{Illustration of the deceleration
 a heavy particle $m$ experiences when
  moving in an homogeneous and infinite ``sea'' of much lighter particles
  $m_*$.} 
\label{fig:fig0}
\end{figure}

Consider a particle $m$ moving with velocity $v_m$ 
in an homogeneous background of stationary lighter
particles of equal mass $m_*$; see
Figure~\ref{fig:fig0}.  Assume only changes in kinetic energy. 
As $m$ moves through, a particle $m_*$ incoming with impact parameter 
$b$ will be given a velocity impulse of about
 the acceleration $a$ times the duration of the encounter $\Delta t$. This can
 be approximated as
\begin{equation}
\Delta v_* \approx \frac{G m}{b^2}\times \frac{b}{v_m} \,.
\end{equation}
The kinetic energy gain of $m_*$ is therefore
\begin{equation}
\Delta E_* \approx \frac{1}{2} m_* (\Delta v_*)^2 \approx  
\frac{1}{2} m_*  \left( \frac{G m}{b v}\right)^2 \,.
\end{equation}

The total change in velocity of the massive particle is given
by accounting for all the encounters it suffers with particles $m_*$. The
number of encounters with impact parameter between $b$ and
$b+\Delta b$ is  $\Delta N \approx n_0\, (v_m \Delta t)\,\Delta(\pi b^2)$; where $n_0$
is the number density of background stars. 
The total change in velocity of $m$ at the expense of the energy lost by  stars is then
\begin{equation}
 \frac{{\rm d} v_m}{{\rm d} t} \approx \frac{1}{m v_m}\int \frac{{\rm d} E_*}{{\rm
 d} t}  {{\rm d} N}  \approx  
 \frac{\pi G^2  \rho_0 m}{v_m^2} \int_{b_{\rm min}}^{b_{\rm max}}
 \frac{ {\rm d} b}{ b} \,,
\end{equation}
where we set $\rho_0=n_0 m_*$, the background density, and  $b_{\rm min}$ and $b_{\rm max}$ are a minimum
and maximum impact parameter, respectively. Letting $\ln \Lambda$ be 
the resulting integral, the 
 deceleration of $m$ due to its
interaction with an homogenous background of particles stars is
\begin{equation}
\frac{{\rm d} v_m}{{\rm d} t} \approx \frac{\pi G^2  \rho_0 m}{v_m^2} \ln \Lambda \,.
\label{eq:df_0}
\end{equation}

The velocity impulse on $m_*$ has a perpendicular $\Delta v_\perp$ and 
parallel $\Delta v_{||}$
component; see Figure~\ref{fig:fig0}. It is not difficult to see that a mean
vector sum of all the $\Delta v_\perp$ contributions  vanishes in this case.
This is not true  however for the mean square of $\Delta
v_\perp$.\footnote{\footnotesize Contributions from $(\Delta
v_\perp)^2$ are linked to the concept of relaxation time in stellar
systems.\cite{BT,Saslaw}} Thus  the dynamical friction force 
is along the line of motion of $m$.

Several key features of dynamical friction are observed from equation
(\ref{eq:df_0}) in this elementary calculation,  that appear also in more
elaborate treatments. (1) The
deceleration of the massive particle is proportional to its mass $m$,  
 so the frictional force it experiences is directly proportional to $m^2$. 
 (2) The
 deceleration is inversely 
 proportional to the square of its velocity $v_m$.

\subsection{Chandrasekhar formula}\label{sec:df2}

A further step in calculating the effect of dynamical friction is to consider
hyperbolic Keplerian two-body encounters. Such analysis was done by
Chandrasekhar.\cite{Chandra1943,ChandraBook} The resulting formula is provided in textbooks on stellar
dynamics.\cite{BT} For completeness such calculations is provided here,
following  Binney \& Tremaine.

\begin{figure}
\includegraphics[width=7.5cm]{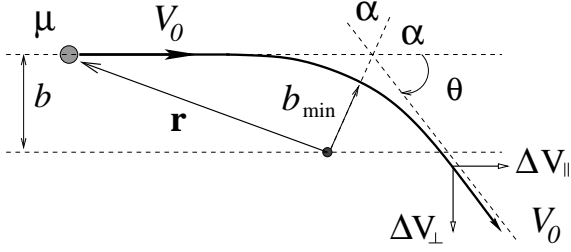}
\caption{Dispersion of a  ``reduced'' mass particle $\mu$
  in the potential of a static body of mass $(m+m_*)$ moving with relative
  speed $V_0$. The scattering angle is $\theta$.}
\label{fig:2body}
\end{figure}

Use of well known results from the
Kepler problem for two bodies in hyperbolic encounters are used.\cite{AF,HRK,BPC1,French,Coffman} 
The two-body problem can be reduced to that of the
motion of a particle of {\sl reduced} mass  $\mu=m m_*/(m+m_*)$ about a fixed
center of force: 
\begin{equation}
\mu {\ddot{\mathbf r}} = - \frac{\kappa}{r^2} {\hat{\mathbf
    r}} \,,
\end{equation}
where $\kappa=G m m_*$, ${\mathbf r}={\mathbf r}_* -{\mathbf r}_m$ is the relative vector position of particles $m$ and
$m_*$, and  ${\hat{\mathbf r}}$ its  unit vector; see
Figure~\ref{fig:2body}. The relative velocity is then
 ${\mathbf V}={\mathbf v}_*-{\mathbf v}_m$, and a change in it is
\begin{equation}
\Delta {\mathbf V}=\Delta {\mathbf v}_*
-{\Delta \mathbf v}_m \,.
\label{eq:cm-1}
\end{equation}

The velocity of the center-of-mass of $m$ and $m_*$ does not change, hence
\begin{equation}
m_* \, \Delta {\mathbf v}_*
+ m \, {\Delta \mathbf v}_m = 0\,.
\label{eq:cm-2}
\end{equation}
From equations (\ref{eq:cm-1}) and (\ref{eq:cm-2}) the change
in velocity of $m$ is
\begin{equation}
 {\Delta \mathbf v}_m  = -(\frac{m_*}{m+m_*})\, \Delta {\mathbf V}\,.
\label{eq:vmassive}
\end{equation}

Once  $\Delta {\mathbf V}$ is determined, ${\Delta \mathbf v}_m$ can be found
from equation (\ref{eq:vmassive}).
From the symmetry of the problem, is better to decompose  $\Delta {\mathbf V}$ in terms of perpendicular and parallel
components:
\begin{equation}
\Delta {\mathbf V} = \Delta {\mathbf V}_{||} + \Delta {\mathbf V}_{\perp}\,,
\label{eq:components0}
\end{equation}
with
\begin{equation}
 |\Delta {V}_{||}| = V_0 \cos \theta  \quad {\textrm{and}}
\quad
| \Delta {V}_{\perp}| = V_0 \sin \theta \,,
\label{eq:components1}
\end{equation}
where $\theta$ is the angle of dispersion and $V_0$ the initial speed at
infinity; this being the same after the encounter since only kinetic energy
changes are considered; see Figure~\ref{fig:2body}. From
geometry, the angle $\alpha$ in Figure~\ref{fig:2body} is related to the orbit's eccentricity $e$ by\cite{Adolph}
\begin{equation}
\cos \alpha = \frac{1}{e} \quad\to\quad \cot \frac{\theta}{2}=\sqrt{e^2-1} \,,
\end{equation}
where $\theta+2\alpha=\pi$. Physically
$e$ is given by
\begin{equation}
 e=\sqrt{1+ \frac{2 E L^2}{\mu^3 \kappa^2}}\,,
\label{eq:ecc0}
\end{equation}
where $E= \mu V_0^2/2$ in the kinetic energy and $L= \mu b V_0$ the angular
momentum magnitude.
Since
\begin{equation}
\sin \theta = \frac{2 \tan \frac{\theta}{2}}{1 + \tan^2 \frac{\theta}{2}}
\qquad {\textrm{and}}\quad
\cos \theta = \frac{1 - \tan^2 \frac{\theta}{2} }{1 + \tan^2 \frac{\theta}{2}}
\,,
\label{eq:trig1}
\end{equation}  
after some algebra  it is found that
\begin{equation}
 |\Delta {V}_{\perp}| = 
\frac{2 b V_0^3}{G(m+m_*)} \left[ 1 + \frac{b^2 V_0^4}{G^2
 (m+m_*)^2} \right]^{-1} \,,
\label{eq:vperp}
\end{equation}
\begin{equation}
 |\Delta {V}_{||}| = 2 V_0 \left[ 1 + \frac{b^2 V_0^4}{G^2
 (m+m_*)^2} \right]^{-1} \,.
\label{eq:vparallel}
\end{equation}

Using equation (\ref{eq:vmassive}) the perpendicular and parallel magnitudes
of the components of $\Delta {\mathbf{v}}_m$ follow:
\begin{equation}
 |\Delta {\mathbf{v}}_{m\perp}| =  \frac{2 b m V_0^3}{G(m+m_*)^2} \left[ 1 + \frac{b^2 V_0^4}{G^2 (m+m_*)^2} \right]^{-1}\,,
\label{eq:vm1}
\end{equation}
\begin{equation}
 |\Delta {\mathbf{v}}_{m||}| =  \frac{2 m V_0}{(m+m_*)} \left[ 1 + \frac{b^2 V_0^4}{G^2
 (m+m_*)^2} \right]^{-1} \,.
\label{eq:vm0}
\end{equation}
In a homogeneous sea of stellar
masses all perpendicular deflections cancel by symmetry.  However,
the parallel velocity changes are added and the mass $m$ will experience a
deceleration. 

The calculation of the total drag force due to a set of particles $m_*$ is
as follows.
Let $f(\mathbf{v}_*)$ be the number density of stars.
The rate at which particle $m$ encounters stars with impact
parameter between $b$ and $b + {\rm d} b$, and velocities between
${\mathbf{v}}_*$ and ${\mathbf{v}}_* + {\rm d}{\mathbf{v}}_*$, is
\begin{equation}
2 \pi b \, {\rm d}b \cdot V_0 \cdot f(\mathbf{v}_*) \, {\rm d}^3 \mathbf{v}_* \,,
\end{equation}
where ${\rm d}^3  \mathbf{v}_*$ is the volume element in velocity
space. The total change in velocity of $m$ is found by adding all the
contributions of $ |\Delta {\mathbf{v}}_{m||}|$ due to particles with impact parameters
from $0$ to a $b_{\rm max}$ and then summing over all velocities of stars. At
a particular  $\mathbf{v}_*$ the change is
\begin{equation}
\frac{{\rm d} {\textbf{v}}_m }{ {\rm d}t}\Big|_{{\mathbf v}_*} =  {\mathbf V}_0 \cdot f(\mathbf{v}_*)
\, {\rm d}^3 \mathbf{v}_* \int_0^{b_{\rm max}}  |\Delta  \mathbf{v}_{m||}| \,
 2 \pi b \, {\rm d}b\,.
\label{eq:rate}
\end{equation}
The required integral is
\begin{eqnarray}
{\cal I}&=& \int_0^{b_{\rm max}} \left[ 1 + \frac{b^2 V_0^4}{G^2
 (m+m_*)^2} \right]^{-1} \, b \, {\rm d}b \nonumber \\
&& = \int_0^{b_{\rm max}} \frac{b \,
{\rm d}b}{1 + a b^2} 
=\frac{1}{2 a} \int_{1}^{s_{\rm max}} \frac{{\rm d}s}{s} \,, \nonumber
\end{eqnarray}
where  $a=V_0^4/G^2(m+m_*)^2$ and $s=1+a b^2$, with $s_{\rm max}=1
+ a b^2_{\rm max}$. Evaluating the integral yields
$$
 {\cal I}= \frac{1}{2}\frac{G^2 (m+m_*)^2}{V_0^4} \ln \left[ 1 + 
\Lambda^2 \right]\,,
$$
where
\begin{equation}
\Lambda \equiv \frac{b_{\rm max} V_0^2}{G (m+m_*)} = \frac{b_{\rm max}}{b_{\rm min}} \,.
\label{eq:lambda}
\end{equation}
Putting these results together in equation (\ref{eq:rate}):
\begin{eqnarray}
\frac{{\rm d} {\textbf{v}}_m }{ {\rm d}t} \Big|_{{\mathbf v}_*} &=& 2 \pi G^2 \ln (1 + \Lambda^2) m_*
(m+m_*) \nonumber \\
& &\times\;  f(\mathbf{v}_*)
\, {\rm d}^3 \mathbf{v}_*  \frac{ {\mathbf{v}}_* -  {\mathbf{v}}_m  }{
  |{\mathbf{v}}_* -  {\mathbf{v}}_m |^3 } \,.
\label{eq:df1particle}
\end{eqnarray}

 The quantity $\ln \Lambda$ is called the {\sl Coulomb logarithm} in analogy
 to an equivalent logarithm found in the theory of plasma. The factor $\ln
 \Lambda$ reflects the fact that the cumulative effect of small deflections
 is more important than strong or close encounters. This may be seen
 geometrically from Figure~\ref{fig:2body}, were the stronger the  deflection
  the smaller is the parallel component contributing to the slow down of $m$.

The determination of the limits
 $b_{\rm min}$ and $b_{\rm max}$ is not  an easy matter and depends on the
 problem at hand. In this approximation $b_{\rm
   min}$ satisfies $V_0^2=Gm/b_{\rm min}$, where $V_0$ depends on the relative
 velocity of $m$ and $m_*$. If the motion of $m$ is relatively slow in
 comparison to that of the stars, $V_0$ can be approximated for example by 
 the root-mean-square value velocity of stars $V_{\rm rms}$. The outer
   limit  $b_{\rm  max}$ is in principle the radius at which stars no longer
   can exchange momentum with $m$. If $m$ is close to the center of a stellar
 system $b_{\rm max}$  can be taken as a particular scale-radius of the
 system; for example, where the star density falls to half of its central value. 

  In typical astronomical applications $\Lambda \gg 1$.
For example, consider the motion of a
massive black hole of mass $m \approx 10^{5} {\rm M}_\odot$ near the center of
a dwarf galaxy.  These galaxies have  $V_{\rm rms}\approx V_0 \approx
30~$km~s$^{-1}$, characteristic radii $b_{\rm max}\approx 3~$kpc and stars
of masses  $m_*\approx 1 {\rm M}_\odot$. Using these values we
obtain $\Lambda \approx 6.3 \times 10^3$. This allows to use 
the approximation $\ln (1 + \Lambda^2) \approx 2 \ln \Lambda$. Note that $\ln
\Lambda$ shows a weak dependence on $V_0$ that is usually neglected. 
 Values of  $2 \lta \ln \Lambda \lta 20$ are typically found in astronomical
 literature.

Now, the integration of equation (\ref{eq:df1particle})
 over the velocity space of stars is required. 
 Writing equation (\ref{eq:df1particle})  as
\begin{eqnarray}
\frac{{\rm d} {\textbf{v}}_m }{ {\rm d}t} &=& G \int_0^\infty 
\frac{  \rho({\mathbf v}_*) ({\mathbf v}_* -  {\mathbf v}_m)   }{ |{\mathbf
    v}_* -  {\mathbf v}_m|^3 }\, {\rm d}^3 {\mathbf v}_* \,,\\
 \rho({\mathbf v}_*) &\equiv& \vphantom{\int} 4 \pi G (m+m_*) m_* \ln \Lambda  f(\mathbf{v}_*)\,,
 \nonumber
\label{eq:rhovm}
\end{eqnarray}
it is noticed that represents the equivalent problem of
 finding the gravitational field (acceleration) 
 at the ``spatial'' point ${\mathbf v}_m$ generated by the ``mass density''
 $\rho({\mathbf v}_*)$. 
From gravitational potential theory,\cite{BT,CollinsCM} 
 the acceleration at a particular spatial point ${\mathbf r}$ is given by
$$
{\textbf{a}}({\bf r}) = G \int_0^\infty
 \frac{ \rho({\mathbf r}') ({\mathbf r}' -
  {\mathbf r})\, {\rm d}^3 {\mathbf r}' }{|{\mathbf r}' -  {\mathbf r}|^3 } =
- \frac{G {\mathbf{r}}}{r^3}  \int_0^r  \rho({\mathbf r}') \, {\rm d}^3
{\mathbf r}'\,.
$$
This is the known result that only matter inside a particular radius
contributes to the force.
In analogy to the gravitational case, the acceleration is given by the
total ``mass'' inside $v_* < v_m$, is
$$
\frac{{\rm d} {\textbf{v}}_m }{ {\rm d}t} = - \frac{G {\mathbf{v}_m}}{v_m^3}
 \int_0^{v_m}  \rho({\mathbf v}_*) \, {\rm d}^3 {\mathbf v}_* \,.
$$
For an isotropic velocity distribution:
\begin{eqnarray}\label{eq:chandra}
\frac{{\rm d} {\textbf{v}}_m }{ {\rm d}t} &=& - {\cal C}_{\rm df}\,
\mathbf{v}_m \,, \\
{\cal C}_{\rm df} &\equiv & 16 \pi^2 G^2 m_* (m+m_*) \ln \Lambda \int_0^{v_m}
f(v_*) v_*^2 \, {\rm d} v_* \,.\nonumber
\end{eqnarray}
This is called {\sl Chandrasekhar dynamical friction formula}. It 
shows that only stars moving slower than $v_m$ contribute to the drag
force on the massive particle.

If stars have a Maxwellian velocity distribution function,
\begin{equation}
f(v_*) = \frac{n_0}{(2 \pi \sigma^2)^{3/2}}\, {\rm e}^{-v_*^2/(2 \sigma^2)}\,,
\label{eq:maxwellian}
\end{equation}
the integral in (\ref{eq:chandra}) in done by an elementary method. In 
dimensionless form it is
$$
{\cal I}_m  = 
\frac{n_0}{\pi^{3/2}} \int_0^{X}  {\rm e}^{-y^2} y^2 \,{\rm d} y \,,
$$
where  $y^2=v_*^2/2\sigma^2$ and $X\equiv
v_m/(\sqrt{2}\sigma)$. Integrating by parts results in 
$$
{\cal I}_m  = \frac{n_0}{4 \pi} \left[ {\rm Erf}(X) - \frac{2 X}{\sqrt{\pi}}\, {\rm
    e}^{-X^2} \right] \,,
$$
where ${\rm Erf}(x)=(2/\sqrt{\pi})\int_0^x {\rm e}^{-y^2} {\rm d} y$ is the
error function.
If  $\rho_0=n_0 m_*$, the density of the background of stars, 
and assume that
$m \gg m_*$, the deceleration of $m$ inside an 
homogeneous stellar system with 
isotropic velocity distribution is:
\begin{eqnarray}\label{eq:df-iso}
\frac{{\rm d} {\textbf{v}}_m }{ {\rm d}t} &=&  - \Gamma_{\rm df} \, \mathbf{v}_
\\
\Gamma_{\rm df} &\equiv &
- \frac{4 \pi G^2 \ln \Lambda \rho_0 m}{v_m^3} \left[ {\rm Erf}(X) - \frac{2 X}{\sqrt{\pi}}\, {\rm
    e}^{-X^2} \right]. \nonumber
\end{eqnarray}

\section{An Analytical Example}\label{sec:analytical}

\begin{figure}[!t]
\includegraphics[width=5cm]{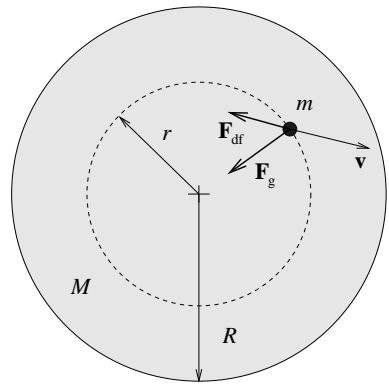}
\caption{Forces acting on a massive particle $m$ moving with velocity
  $\mathbf{v}$ inside a stellar system: $\mathbf{F}_{\rm g}$ is the
  gravitational force and $\mathbf{F}_{\rm df}$ the dynamical friction force. 
 The latter acting opposite to the direction of motion of $m$.}
\label{fig:df2}
\end{figure}

A simple application of Chandrasekhar's formula 
(\ref{eq:df-iso}) for an homogeneous spherically symmetric stellar
 system, although not
infinite, is presented. The problem consists in determining the motion of a 
massive particle $m$ subject to  gravitational and dynamical friction forces.
The stellar system has a radius $R$ and total mass $M$; see
Figure~\ref{fig:df2}. 
The  equation of motion for $m$ is
\begin{equation}
m \frac{ {\rm d}^2 {\mathbf{r}} }{ {\rm d} t^2 } = {\mathbf{F}}_{\rm g} + 
{\mathbf{F}}_{\rm df} = - m \nabla \varphi(r)  + m
\, {\mathbf{a}}_{\rm df}\,,
\label{eq:motion}
\end{equation}
 where $\varphi(r)$ is the gravitational potential, and $\mathbf{a}_{\rm df}$
 is given by equation (\ref{eq:df-iso}).

Equation~(\ref{eq:motion}) is not in general 
tractable by analytical methods, so some approximations are required. 
 Zhao\cite{Zhao04} has found an approximation to the
term associated with the velocity distribution in equation
(\ref{eq:df-iso}), namely:
\begin{equation}
\chi(X) \equiv \frac{1}{X^3}
\left[ {\rm{Erf}}(X) - \frac{2 X}{\sqrt{\pi}} {\rm e}^{-X^2}\right]  \approx
\frac{1}{\frac{4}{3} + X^3}\,,
\label{eq:zhao}
\end{equation}
that works to within 10 percent for $0 \le X < \infty$. When $m$
moves slow in comparison to the velocity dispersion of stars, $v\ll \sigma$,
 $\chi(0) \approx 3/4$, and when $v= \sigma/2\,$  
 $\chi(1) \approx 3/7$. Note that in the case of a very fast
relative motion of $m$  dynamical friction is negligible; a situation analogous
to when a block of material slides fast. Using the previous approximation, and
considering $\chi= 3/4$, the frictional force
$\mathbf{F}_{\rm df}$ in equation (\ref{eq:motion}) becomes: 
\begin{equation}
{\mathbf{F}}_{\rm df} = 
m {\mathbf{a}}_{\rm df} \approx 
- \frac{3 \pi G^2 \ln \Lambda \rho_0 m^2}{(\sqrt{2} \sigma)^3} \,\mathbf{v}
 = - \gamma  \,\mathbf{v}
\,,
\label{eq:df_2}
\end{equation}
where $\gamma = 3 \pi G^2 \ln \Lambda \rho_0 m^2/ (\sqrt{2} \sigma)^{3}$.

To determine $\mathbf{F}_{\rm g}$  recall that
 the potential is related to the density through Poisson equation,
\begin{equation}
\nabla^2 \varphi(r) =  4 \pi G \rho(r) \,,
\end{equation}
whose solution for a spherically symmetrical system of radius $R$ is
\begin{equation}
\varphi(r) = - 4 \pi G \left[ \frac{1}{r} \int_0^r \rho(r) r^2 \, {\rm d} r
+ \int_r^R \rho(r) r \, {\rm d} r \right]\,.
\label{eq:potential}
\end{equation}
In a constant density $\rho_0$ system the potential is 
\begin{equation}
\varphi(r) = -2 \pi G \rho_0\, ( R^2 - \frac{1}{3} r^2 ) \,,
\label{eq:pot}
\end{equation}
and the gravitational force on $m$ is
\begin{equation}
{\mathbf{F}}_{\rm g} =  - m \nabla \varphi(r) = - \frac{4}{3} \pi G \rho_0 m \,
{\mathbf{r}} = - k\, {\mathbf{r}} \,,
\label{eq:sho}
\end{equation}
with $k=4 \pi G \rho_0 m/3$.  This is the well known result
 from introductory mechanics that a particle inside an homogeneous gravitational
 system performs a harmonic motion.

Combining equations (\ref{eq:df_2}) and (\ref{eq:sho}) the resulting equation
of motion is
\begin{equation}
m \frac{ {\rm d}^2 {\mathbf{r}} }{ {\rm d} t^2 } +  k\, {\mathbf{r}} 
 + \gamma  \,\mathbf{v} = 0 \,.
\label{eq:damped}
\end{equation} 
This is the same equation, for example,
 as that of a mass attached to a spring with stiffness constant $k$
 inside a medium of viscosity $\gamma$; 
i.e., a damped
 harmonic oscillator.\cite{AF,HRK,BPC1,French} The solution of equation (\ref{eq:damped}) in a plane,  under arbitrary
initial conditions
$$
x(0)=x_0, \quad {\dot x}(0)=u_0\,; \quad \quad
y(0)=y_0, \quad {\dot y}(0)=v_0 \,,
$$
where the dot indicates a time derivative, is\cite{Weinstock,Luthar}
\begin{eqnarray}
x(t) &=& \frac{{\rm e}^{-(\beta+\xi)t} }{2 R} \left\{ 
( {\rm e}^{2\xi t}-1) u_0 +  \nonumber \right. \\ 
&& \left. \left[ ( {\rm e}^{2\xi t}-1) \beta +
 ( {\rm e}^{2\xi t}+1) \xi \right] x_0 
\right\}\,, \nonumber
\end{eqnarray}
\begin{eqnarray}
y(t) &=& \frac{{\rm e}^{-(\beta+\xi)t} }{2 \xi} \left\{ 
( {\rm e}^{2\xi t}-1) v_0 +  \nonumber \right. \\ 
&& \left. \left[ ( {\rm e}^{2\xi t}-1) \beta +
 ( {\rm e}^{2\xi t}+1) \xi \right] y_0 
\right\}\,,
\label{eq:motionXY1}
\end{eqnarray}
where $2\beta=\gamma/m$ and $\xi=\sqrt{\beta^2-\omega_0^2}$, with $\omega_0^2=
k/m$.

The behavior of $m$ is dictated by the relative values of $\beta$ and
$\omega$. 
The values of $\ln \Lambda$ and $\sigma$ are first to be estimated. Take $b_{\rm max} \approx R$ and $b_{\rm min}= G m/V_0^2$. An
estimate of $V_0$ may be obtained from the virial theorem,\cite{AF}
 that relates 
the kinetic $T$ and potential energy $W$ of the system by:
\begin{equation}
2 T = - W \,.
\label{eq:virial}
\end{equation}
For an homogeneous system of size $R$ the  potential energy is
\begin{equation}
W = - 4 \pi G \int_0^R \rho_0 M r \, {\rm d} r = - \frac{3}{5} \frac{G
  M^2}{R}\,, 
\end{equation}
and the kinetic energy is taken as $T=M V_0^2/2$. This leads to
\begin{equation}
V_0^2 \approx \frac{3}{5} \frac{G M}{R} \approx 3 \sigma^2 \,;
\label{eq:vel}
\end{equation}
where the last term provides an estimate of the one-dimensional velocity
dispersion under the assumption of isotropy in the velocity distribution of
stars. Using equation (\ref{eq:vel}) $\beta$ and $\omega$ are:
\begin{equation}
\beta= \frac{45}{16}\sqrt{\frac{5}{2}} \frac{G^2 m M\ln \Lambda}{R^3
  (GM/R)^{3/2}}\,, \quad\quad
\omega_0 = \sqrt{ \frac{G M}{R^2}}\,.
\label{eq:betaomega0}
\end{equation}
The resulting Coulomb logarithm is $\ln \Lambda = \ln [3 M/(5m)]$.

To compare the numerical values of $\beta$ and $\omega_0$ is better to use
another system of units than a physical one. Let  
$G\!=\!M\!=\!R\!=\!1$, that is a common choice in $N$-body simulations 
in astronomy; 
to return to physical units one can use Newton's law and set $G$ to
the appropriate value (see Appendix).  In these units,  
relations (\ref{eq:betaomega0}) become
\begin{equation}
\beta = \frac{45}{16}\sqrt{\frac{5}{2}} m \ln \left(\frac{3}{5m}\right)
 \,, \quad \quad
\omega_0 = 1 \,. 
\end{equation}
If  $m=1/100$ then $\ln \Lambda \approx 4$ and
  $\beta\approx 0.2 < \omega_0$. Hence an underdamped harmonic motion for the
  massive particle results. If $m=1/10$ 
then $\ln \Lambda \approx 2$ and  
$\beta \approx 0.8\approx \omega_0$, so the motion
of $m$ will be strongly damped. Note that an upper limit to $m$
 is set when $m=3/5$, leading to $\ln \Lambda = 0$; i.e.,  no
dynamical friction results. For larger $m$ a negative $\beta$ is
obtained. Clearly, the model fails and the
behavior of the dynamics is unrealistic.

For cases of interest, where  $m \ll M$, it follows that 
$\beta < \omega_0$ and
the resulting motion (\ref{eq:motionXY1}), after some algebra, is
\begin{eqnarray}
x(t) &=& \left[ x_0 \cos \omega t + \frac{u_0 + \beta x_0}{\omega} \sin \omega
t \right] {\rm e}^{-\beta t} \,, \quad \nonumber \\
y(t) &=& \left[ y_0 \cos \omega t + \frac{v_0 + \beta y_0}{\omega} \sin \omega
t \right] {\rm e}^{-\beta t} \,,
\label{eq:analytic}
\end{eqnarray}
where $\omega^2= \omega_0^2 - \beta^2$. Note that a time-scale when the
orbit decays $1/{\rm e}$ is given by $\tau_{\rm df}=1/\beta$.

\begin{figure}[!t]
\includegraphics[width=8.5cm]{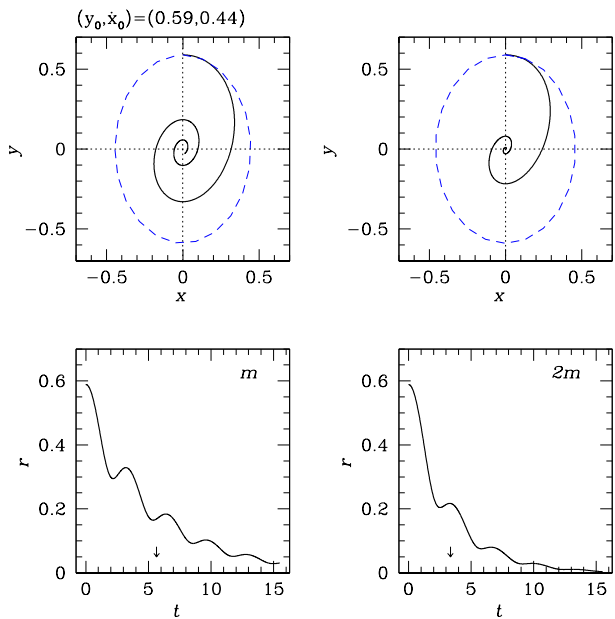}
\vspace{-0.5cm}
\caption{Orbital decay of a massive particle $m$ in a homogeneous
 stellar system due to
  dynamical friction. Initial conditions are $(x_0,y_0)=(0,0.59)$ and
  $(u_0,v_0)=(0.44,0)$. Left panels are for $m=0.01$ and right ones
  for $m=0.02$. Top panels show the orbit (solid line) in the xy-plane and
  bottom ones the time evolution of the distance $r$ from the center. The
  dynamical friction time scale $\tau_{\rm df} = 1/\beta$ is indicated by an
  arrow.  Dashed lines are orbits without considering dynamical friction.} 
\label{fig:orbit0}
\end{figure}

Take as example $m=0.01$. If the
initial position of $m$ is at $(x,y)=(0.59,0.0)$ and its velocity $(0.0,0.44)$
 the resulting orbit is that shown in  
Figure~\ref{fig:orbit0}-{\it
  left} with solid line. Doubling $m$ results in the orbit shown in
 Figure~\ref{fig:orbit0}-{\it right}. The dashed lines in
 correspond to the orbit of $m$ without dynamical
friction.  The
  effect of increasing the mass of $m$ on the orbit,
  and on the decay time $\tau_{\rm df}$, is clearly appreciated.

This example shows the basic features of, for example, the orbital decay of
a satellite galaxy toward the center of its host larger galaxy. It may
be applied also to the motion of a massive black hole near the center of a
galaxy or star cluster, where to some approximation the gravitational
potential can be taken as harmonic. More realistic situations require however
the numerical integration of the orbit and/or an $N$-body computer
simulation. A particular case of these are treated next.

\section{A more realistic example}\label{sec:realistic}

 Chandrasekhar's formula (\ref{eq:df-iso}) although derived assuming an
 infinite homogeneous system may be applied, to some degree, when  stellar
 systems are non-homogeneous.\cite{BT} In this case, 
  local values for the density  $\rho(r)$ and the velocity dispersion
 $\sigma(r)$ are used. Here the motion of a massive particle $m$
 inside a non-homogeneous  stellar system is considered, both using a
 semi-analytical method and $N$-body simulation, to illustrate further the application of dynamical friction.

\subsection{Semi-analytic treatment}\label{sec:semi}

A simple representation of a stellar system, such as a globular cluster or an
elliptical galaxy, is provided by the 
Plummer model. Its potential and stellar density are, respectively:\cite{BT,Saslaw} 
\begin{equation}
\varphi(r) = - \frac{G M}{(r^2 + a^2)^{1/2}}, \quad 
\rho(r) = \frac{3 M a^2/4 \pi}{(r^2 +a^2)^{5/2}},
\label{eq:plummer}
\end{equation}
where $M$ is the total mass, and $a$ the scale-radius of the system. 
In a spherical system with isotropic velocity distribution the 
equation of ``hydrostatic'' equilibrium\footnote{\footnotesize In mechanical
  equilibrium a change in pressure ${\rm d}P$ is balanced by the gravitational
``force'' $-\rho(r)\nabla \varphi(r)\,{\rm d}r$. The pressure is here $P=\rho
\sigma^2$, similar to that of an ideal gas where $P=kT\rho/m$. The equation
used is a particular case of that called in stellar dynamics Jeans equation.} is satisfied:
\begin{equation}
\frac{1}{\rho} \frac{{\rm d} (\rho \sigma^2)}{{\rm d} r} = - \frac{{\rm d}
  \varphi}{{\rm d} r}  \quad \rightarrow \quad
\sigma^2(r) = -\frac{\varphi(r)}{6}\,.
\label{eq:Jeans}
\end{equation}
The last result follows from noticing that
 $\rho \propto \varphi^5$, and imposing boundary 
conditions that both $\rho \sigma^2$ and $\varphi$  go to zero at
 infinity. 

Equations (\ref{eq:plummer}) and (\ref{eq:Jeans}) will be used in 
 equation  (\ref{eq:df-iso}) to compute the orbital motion of a massive particle $m$. It rests to determine $b_{\rm min}$ and $b_{\rm max}$. 
The former is evaluated at local values,
 $b_{\rm min}\!=\!Gm/[3\sigma^2(r)]$, and the latter is set fix to $b_{\rm max}\!=\!a$.

The equation of motion (\ref{eq:motion})  for $m$ can now be integrated
numerically using standard methods,\cite{NR,Garcia} or using the one discussed
 by Feynman\cite{Feynman1} for planetary orbits ($\S$9). Here 
a fourth-order  Runge-Kutta algorithm with adaptive time-step was used. 
The initial
conditions for $m$ are the same as those used in the analytical case.

In Figure~\ref{fig:plumDF} the resulting orbit from the numerical integration 
is shown as a dashed line. Also, the behavior of the $x$ and $y$ coordinates,
and of the distance $r$ of $m$ to the center, as a function of time are shown.
The typical decay of the orbit is evident.
 In the same figure results from an $N$-body simulation are displayed, that are
described next.

\subsection{N-body simulation}\label{sec:nbody}

The use of $N$-body simulations allows to study more realistically the different
dynamical phenomena that occur in stellar systems.\cite{Hockney,Aarseth} 
Several $N$-body codes with different degrees of sophistication have been
developed for astronomical problems in mind.\cite{BH,gadget2,Dehnen} 
 Some low-$N$ simulations can be run nowadays using a personal computer with publicly available $N$-body codes.\footnote{\footnotesize The reader may
  obtain, for example, Barnes'  {\sc tree-code} at
  http://www.ifa.hawaii.edu/faculty/barnes/software.html. The site contains
  also programs, both in ${\sc C}$ and {\sc Fortran}, to generate some stellar
  systems and initial conditions. The {\sc Gadget} code of Springel is
  at http://www.mpa-garching.mpg.de/gadget/. Dehnen's {tree-code} is included
  in the {\sc Nemo} package under {\sc gyrfalcON} at
  http://bima.astro.umd.edu/nemo/.}

 Barnes' {\sc tree-code} in {\sc Fortran}, and some of his  public
 subroutines are used to simulate the motion of  $m$ 
inside a Plummer model. 
A numerical realization of this model with $N=10^5$ particles is used 
with individual ``star'' masses of $m_*=1/N$. 
 The massive particle $m=1/100$ with initial conditions
 $(y_0,{\dot x}_0)=(0.59,0.44)$ is set ``by hand'' inside the
 numerical Plummer model.  In $N$-body units the scale radius is
 $a=3\pi/16=0.59$ (see Appendix). 

The circular period at radius $r$  is
 $\tau=2\pi r/V_{\rm c}(r)$, where the circular velocity and integrated mass
 for a Plummer model are given, respectively, by:
$$
V_{\rm c}(r)=\sqrt{\frac{GM(r)}{r}}\,, \quad \quad
M(r) = \frac{M \,(r/a)}{[1+(r/a)^2]^{3/2}}\,.
$$
From this, an orbital period of $\tau_{\rm
  a}=4.8$ time units at $r=a$ results. 
The simulation was run for $t=10\approx 2\tau_{\rm a}$ time units.
The parameters for running the 
 {\sc tree-code} in serial were those provided by Barnes
 at his Internet site for an
 isolated  Plummer evolution. The quadrupole moment in the
 gravitational potential is activated.
 The simulation took
 about 5.2~{\sc cpu} hours on a PC with an Athlon 2.2GHz
 processor, and 512~KB of cache size. Energy conservation was $\le 0.04$
 percent, that is considered very good.

\begin{figure}
\includegraphics[width=8.5cm]{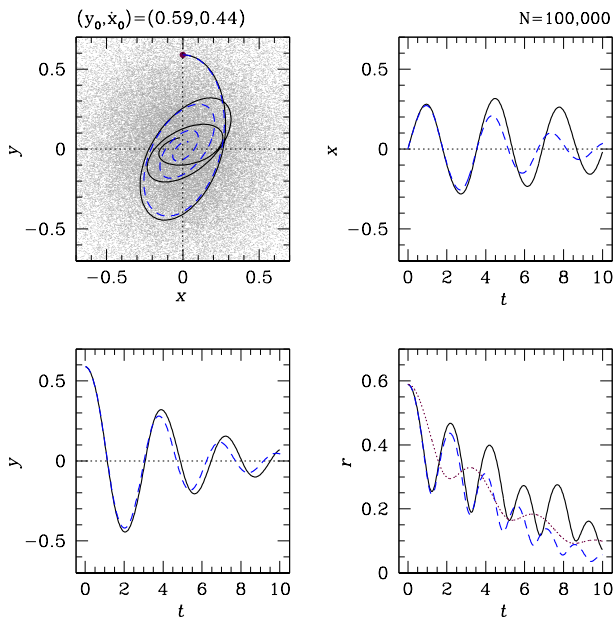}
\vspace{-0.5cm}
\caption{$N$-body simulation of the orbital decay ({\it solid line}) of a
  massive 
  particle $m$ inside a Plummer stellar model. 
 The semi-analytical ({\it dashed line}) 
  and the analytical calculation ({\it dotted line}) of
  Figure~\ref{fig:orbit0} are drawn for comparison. These
  overestimate the effect of dynamical friction in comparison to the
  numerical simulation.} 
\label{fig:plumDF}
\end{figure}

Figure~\ref{fig:plumDF} shows the orbital evolution of the massive particle
$m$ in the $N$-body system as a solid line. The dashed line corresponds to the
semi-analytical calculation of Section~\ref{sec:semi}. This follows closely
the orbit of $m$ in the $N$-body simulation for  about $\tau_{\rm a}$ time units.  
Afterwards, it deviates from the $N$-body result. In the $r$-$t$  panel, 
 the analytical solution (\ref{eq:analytic}) is
shown as a dotted line; that is,  assuming 
the total system was homogeneous. 

Both approximations overestimate
the decay rate of $m$ in comparison to the $N$-body simulation. 
Taking $b_{\rm max}=a/5$ leads to a somewhat better agreement, but
does not reproduce the $N$-body result. Rather surprisingly, the analytical
result does a fair job in reproducing the overall orbital decay in this case.

\section{Final Comments}

The approximations in deriving Chandrasekhar formula limits, obviously, its
application to more complex stellar systems than the one considered here. 
However, it is remarkable  that
equation (\ref{eq:df1particle}) leads to reasonably well results when used with
values under a local approximation.  

In similar vain to the study of the friction between
surfaces,\cite{FrictionRL01,FrictionRL02,Ringlein} dynamical friction is a
complex subject. Elaborate calculations based on Brownian motion,\cite{Chandra49} linear response theory, resonances, and the fluctuation-dissipation theorem 
exist.\cite{BM92,W89,NT99} These that are 
steps forward toward a more complete physical theory for this process.

Instead of listing explicitly some of the shortcomings of Chandrasekhar
dynamical friction formula\cite{BT} when applied to gravitational systems, 
the student is encourage to think on some of 
them and possible improvements on such formula.

From the point of view of an introductory or intermediate class on mechanics
the exposure of students to non-typical problems, as the one presented here 
contributes to further their understanding and appreciation of the subject.

Some ideas that may lead to  problems and/or
 projects for students are:
\begin{enumerate}

\item  How would the analytical solution considered here would be changed if
  the Plummer model is used? What type of approximations would be required to
 make? How does $\ln \Lambda$ change?

 \item If $\sigma^2$ is a measure of the kinetic energy per unit mass of stars,
 what is an
 estimate for its  mean increase due to the energy lost by the massive particle
 during its decay? 

\item How would the orbital decay time be changed
 for different types of initial eccentricities of the massive particle?

\item  Consider a star cluster ($m=10^6 \,{\rm M}_\odot$) in circular orbit 
  at a distance of $r=5\,{\rm kpc}$ from the center of our
 galaxy ($M\approx 6\times 10^{11} {\rm M}_\odot$, $R\approx 150\,$kpc). Would it be expected to fall to the
 center within the age of the universe, say  $t=10^{10}\,$yr? Typical
 velocities for stars and dark matter particles
 at that distance are about $200$~km/s, and the
 scale-radius may be around 5~kpc. What if instead of
 a star cluster we have a galaxy satellite, such as the Magellanic Clouds, with
 $m\approx 10^{10}\, {\rm M}_\odot$ and at a distance of $100\,$kpc?

\item  How do results change if
 instead of a Plummer model a more pronounced density profile is used, such as
 the Hernquist\cite{Hernquist} model? How does the number of particles $N$
  in a simulation affect the decay rate?

\item As the massive
 particle moves through the stellar system it induces a density wake behind it.  Can this be
 detected in an $N$-body simulation on a home computer? How about looking for this wake in the
 phase-space diagram (e.g. a plot of ${\dot x}$--$x$) of stars near the
 the massive particle? 

\item How good do the local approximation works if instead of a massive particle one has an
 extended object, small in comparison to its host galaxy?
\end{enumerate}

Textbook problems
 are designed in general to yield one correct answer, the above ideas for
problems are rather vague but this is on purpose.
The reason is twofold. On one hand, to promote
in students a spirit of research by setting an approximate physical model and
to look for the required data and ``tools'' to solve it; some of them can be found in the references. 
On the other hand, no single definite answer can be given. A feature proper of the
way physics evolves toward  describing and understanding nature.


\appendix

\section{Astronomical and N-body Units}

Several quantities in astronomy are so large in comparison to common
``terrestial'' values, that special units are used. Table~\ref{tab:table1}
lists some of these and their equivalences in physical units.

\begin{table}
\caption{\label{tab:table1}Astronomical units}
\begin{ruledtabular}
\begin{tabular}{ll}
Unit  & Equivalence \\
\hline
Astronomical unit\footnotemark[1] & ${\rm{AU}}=1.496\times 10^{11}\,$m\\
Parsec    & pc=$\,2.063 \times 10^5\,$AU\\
          & \hphantom{pc}=$\,3.261$ light-years\\
Kiloparsec & kpc=$10^3\,$ pc \\
Solar mass & M$_\odot = 1.989 \times 10^{30}\,$kg \\ 
Year       & yr = $3.156\times 10^7\,$s \\
\end{tabular}
\end{ruledtabular}
\footnotetext[1]{Mean sun-earth distance}
\end{table}

In the {\sc mks} system of units the Gravitational constant is
$G=6.67\times 10^{-11}\,$m$^3$~kg$^{-1}$~s$^{-2}$. 
A natural system of units for gravitational interactions is that where
the gravitational constant is set to  $G=1$; in the same way as for quantum
systems Planck's constant is usually set to $\hbar=1$. 
On dimensional grounds
$
[G] = u_v^2 u_l/u_m$;
where $u_m$, $u_l$, and $u_v$ correspond, respectively,
 to units of mass, length and
velocity. 

The Gravitational constant can be expressed in terms of typical 
astronomical values, for example, as:
$$
G = 4.3007 \times 10^{-3} \frac{{\rm km}^2 {\rm pc}}{{\rm s}^2\,{\rm M}_\odot}
 =  4.4984  \times 10^{-3} \frac{{\rm pc}^3}{{\rm Myr}\,{\rm M}_\odot} \,.
$$
The transformation of $G$ using length units such as kpc or Mpc ($10^6\,$pc) is direct.
 Choosing $u_l$ and $u_m$ the unit of
velocity and of time $u_t$, under an appropriate $G$ value, are
$$
u_v=\sqrt{\frac{G u_m}{u_l}}\,, \qquad
u_t=\sqrt{\frac{u_l^3}{G u_m}}\,.
$$

In this way the transformation from 
$N$-body units, where $G=M=R=1$, to physical ones can be made.
Table~\ref{tab:table2} lists some 
values for different choices of $u_l$ and $u_m$, and the
resulting units of $u_v$ and $u_t$. The entries correspond to using the
approximate size and mass of a globular cluster, a disk of a spiral galaxy,
and of a cluster of galaxies, respectively, as units $u_l$ and $u_m$.

\begin{table}
\caption{\label{tab:table2}From $N$-body units to astronomical}
\begin{ruledtabular}
\begin{tabular}{llrrr}
Stellar system &$u_l$ & $u_m$  & $u_v$  & $u_t$ \\
               &      & M$_\odot$ & km/s & Myr \\
\hline
Globular cluster    & 50~pc & $10^{6\hphantom{1}}$ &  9.3 & 5.3 \\
Galaxy              & 10~kpc & $10^{11}$ &  207.4 & 47.2 \\
Cluster of galaxies & $\;\;5$~Mpc & $10^{15}$ &   927.4 & 5271.4 \\
\end{tabular}
\end{ruledtabular}
\end{table}

In the standardized gravitational $N$-body units\cite{Heggie85,Aarseth} the total energy of a system
is $E=-1/4$. This follows from the virial theorem ($2T+W=0$), where
$$
W = -\frac{1}{2} \frac{GM}{R} \quad \to \quad
E= \frac{W}{2} = \frac{GM}{4 R}.
$$
Here $R$ is strictly what is called the virial radius of the system; that does not
necessarily coincides with the total extent of the stellar system, but is a
very good approximation. 
The
potential energy of a Plummer model is
$$
W= \frac{1}{2}\int_0^\infty \rho(r) \varphi(r) 4 \pi r^2\, {\rm d}r
= - \frac{3 \pi}{32} \frac{GM^2}{a}\,.
$$
Thus the total energy is $E=-(3\pi GM^2)/(64 a)$. In $N$-body units this leads 
to a value of the Plummer scale-radius of $a=3\pi/16$. 

\newpage 

\bibliography{Aceves}
\end{document}